\documentclass{raa}            

\usepackage{graphicx,times}             
\usepackage{natbib}
\usepackage{amssymb,amsmath}
\bibpunct{(}{)}{;}{a}{}{,}

\usepackage[a4paper=true,dvipdfm=true,pagebackref=true]{hyperref}
\hypersetup{colorlinks = true, linkcolor = green, anchorcolor = red, citecolor = blue, filecolor = red, pagecolor = red, urlcolor = red}

\begin{document}

 \title{Simultaneous photometric and spectral analysis of a new outburst of  V1686 Cyg}


   \volnopage{Vol.0 (20xx) No.0, 000--000}      
   \setcounter{page}{1}          

   \author{H.R. Andreasyan
      \inst{}
   \and T.Yu. Magakian
      \inst{}
   \and T.A. Movsessian
      \inst{}
   }

   \institute{Byurakan Observatory NAS Armenia, Byurakan, Aragatsotn prov., 0213, Armenia; {\it hasmik.andreasyan@gmail.com}}
\abstract{We present an analysis of the optical observations of Herbig AeBe star V1686 Cyg, which is associated with a small isolated star-forming region around HAeBe star  BD+40$^{\circ}$4124. We observed this star as a part of  our project of young eruptive stars investigation. Observations were held on 2.6m telescope of Byurakan Observatory from 2015 to 2017. For this period we obtained V1686 Cyg direct images and 14 medium- and low-resolution spectra. In the course of observations we noticed that this star underwent a not-typical brightness outburst. After data reduction we found that the full rise and decline of V1686 Cyg brightness had almost 3 magnitudes amplitude and lasted about 3 months. We were also able to trace the changes of the stellar spectrum during the outburst, which  are correlated with the photometric variations.
 }

   \authorrunning{H.R. Andreasyan et al.}            
   \titlerunning{New outburst of V1686 Cyg}  

   \maketitle

%
%
\section{Introduction}           
\label{sect:intro}

Herbig AeBe stars represent the more massive (M$\geq$2$M_\odot$) class of pre-main sequence (PMS) stars.  These stars are known to demonstrate the different types of variability. Studies of these regular and irregular variations in brightness and spectral appearance have a great importance for better understanding the stellar envelopes and the circumstellar disks surrounding these stars. Based on the features of variability (duration, spectral characteristics, amplitudes of brightening), different classification of these stars exist  (see \citet{Herbig1977}). The reasons for the difficulties in the classification of Herbig AeBe (HAeBe) stars lie mainly in the insufficiency of simultaneous photometric and spectroscopic data. This is why we decided to start the project on the investigation of PMS stars showing  periodic or non-periodic eruptive activity. Among them we studied a group of HAeBe stars in the vicinity of the bright BD+40$^{\circ}$4124 star, which itself is Herbig Be star with very strong emission H$\alpha$ line. One of them is LkH$\alpha$~224 which also is HAeBe star. Along with nearby LkH$\alpha$~225 all three stars for the first time were mentioned by \citet{Herbig} in his famous paper about Ae/Be stars, connected with bright nebulosities. With several fainter  stars they create a small young cluster, embedded in the dense molecular cloud. \citet{Shevchenko1991} estimated a distance to this group as 980 pc. 

The photometric variability of LkH$\alpha$~224 was detected by \citet{Wenzel1980}; as the variable this star received V1686~Cyg designation. The most complete information about its photometric behavior so far was collected in the works of \citet{Shevchenko1991,Shevchenko1993} and \citet{Herbst}. According to these data, the star fluctuates near the mean brightness level,  sometimes (usually one time per year) demonstrating irregular Algol-like minima, which lasts one-two months: steep and rapid drop by 2--2.5 mag  and afterwards more slow and oscillating raise in brightness to mean value. We can take into account that many Herbig AeBe stars with the A0 or later spectral type show this kind of variability \citep{Semkov&Peneva}.  Besides, also long-term variations exist:  mean brightness of V1686~Cyg in V decreased in about 8 years by $\approx$2.5 mag, achieving minimum in 1993 and in subsequent 4 years restored  to previous level; thus, its mean level of brightness itself varies  from V=12.5 down to V=15 \citep{Hillenbrand, Herbst, Oudmaijer}.

V1686~Cyg was the subject of many spectral studies, but  prominent variations of absorption features make its spectral classification very problematic. Estimates of its spectral type  vary from B2 to F9 \citep[see][and references therein]{Hernandez}. As it is stated in the same work, shorter term photometric variations of  V1686~Cyg could be related to the spectroscopic variability, but this question is not investigated yet. It should be noted that emission lines in the V1686 Cyg spectrum also demonstrate significant variability \citep{Mora}. In the  work of \citet{Magakian-a} the broad  asymmetric profile of H$\alpha$  emission with several superposed narrow absorption-like features is shown, suggesting the existence of expanding envelopes.

 In 2015 we started systematic observations of BD+40$^{\circ}$4124 field, since the new outburst of V1318 Cyg S was detected \citep{MMAG}. In parallel to these studies the V1686 Cyg star also was observed photometrically and spectrally. The unusual and unsuspected brightening of V1686 Cyg up to almost 3 mag was found and traced. This paper describes the results of these observations. 

\begin{table}
\caption{Log of the V1686 Cyg spectral observations.}
\centering
\begin{tabular}{c c c c}
\hline\hline
Date & Spec. range (\AA) & Resol. ($\lambda/\Delta\lambda$) & Total exp. (min)  \\
\hline%
22.09.2015 & 5880-6740 & 2500 & 60  \\   
18.11.2015 & 5785-7315 & 1500 & 60  \\
22.11.2015 & 5785-7315 & 1500 & 60  \\
16.05.2016 & 5880-6880 & 2500 & 30  \\
10.06.2016 & 5890-6795 & 2500 & 40  \\
08.08.2016 & 5780-7300 & 1500 & 30  \\
23.08.2016 & 4120-6810 & 800 & 60  \\
24.08.2016 & 5895-6795 & 2500 & 40  \\
29.08.2016 & 4070-7055 & 800 & 20  \\
30.08.2016 & 5870-6875 & 2500 & 40  \\
06.11.2016 & 5695-7360 & 1500 & 45  \\
28.11.2016 & 4025-6995 & 800 & 30  \\
20.12.2016 & 4025-6995 & 800 & 45  \\
21.12.2016 & 5850-6850 & 2500 & 30  \\

\hline\hline
\end{tabular}
\label{spectra}
\end{table}


\section{Observations and data reduction}
\label{sect:Obs}

In 2015 we started our project on investigation of eruptive stars. Observations were carried out on 2.6-m telescope in Byurakan Observatory. We used SCORPIO  spectral camera, and obtained direct images, as well as long-slit spectra \citep{Scorpio}. It was equipped with TK SI-003A $1044 \times 1044$ CCD matrix and after the 2016 August with E2V CCD42-40 $2080 \times 2048$ CCD matrix. As an object of our previous interests we started our observations with V1318 Cyg young variable star, and at the same time obtained data for the neighbor V1686 Cyg star. Observations were implemented from September 2015 to July 2017. 
Data reduction was done in the usual way, using IRAF and ESO-MIDAS programs. Photometric estimations were done by aperture photometry. For calibration we used some stars in the field, while for comparison stars we used data from \citet{Hillenbrand} and \citet{Shevchenko1991}.
The typical errors for our data are about 0.$^{m}$02-0.$^{m}$03. 
We also obtained spectra for this star from Sept. 2015 to Dec. 2016, with different spectral resolutions: 0.50, 0.80, 1.50 and 2.65 A/pix.

 More details about the methods of observation and data reduction can be found in our previous paper \citep{MMAG}. Here, in Table~\ref{spectra}, we present the log of the spectral observations.

\section{Results}
\label{sect:data}
\subsection {The new outburst of V1686~Cyg~}

The analysis of the historical light curve of V1686 Cyg (based on the encompassing eleven years photometric database: \citealt{Herbst}) with VizieR service shows, that from the mean level of brightness this star time to time decrease its brightness by 1--1.5 mag. amplitudes, which last 10-20 days and then it returns to mean value.

\begin{figure}[ht!]
  \centering
  \includegraphics[width=0.50\textwidth]{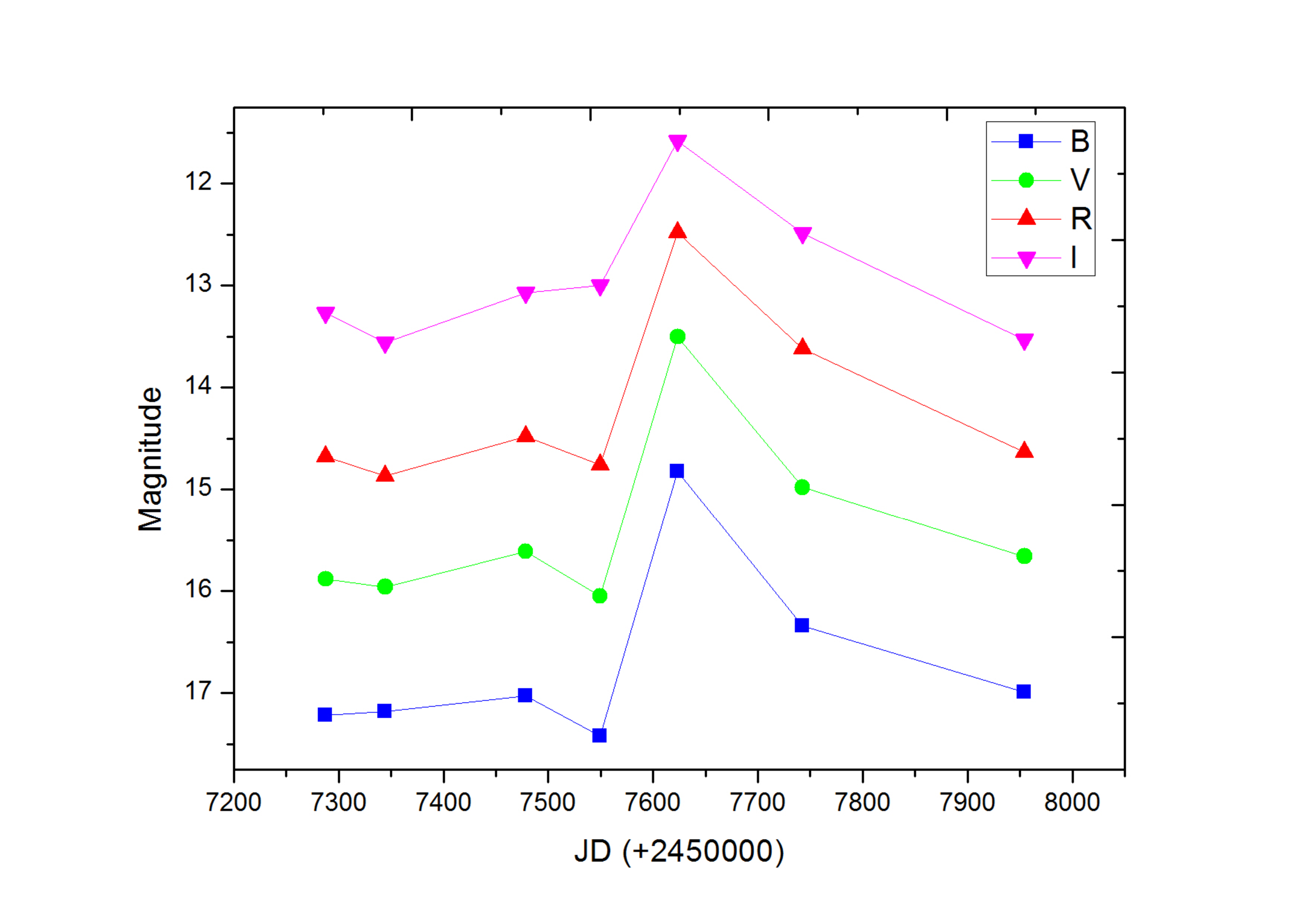}
  \caption{\textit{BVRI} light curve of V1686~Cyg~ for the period of 2015-2017.  \textit{} }
   \label{lightcurve}
\end{figure}

In our case we had the contrary behavior: during the period from Sept.2015 to Aug 2016, i.e. almost 1 year of regular observations any significant photometric variability of V1686 Cyg was not detected. But in Aug. 2016 we noticed that the star significantly raised in brightness. Photometric estimations demonstrated that in that period, which lasted probably several months, the brightness of  V1686 Cyg unsuspectingly increased for more than two magnitudes in V and then  gradually returned to its previous level. Table \ref{phot1}  presents the results of our BVRI photometry, and the light curve is shown in Fig. \ref{lightcurve}. 

To trace the recent variations of V1686 Cyg we checked such sources as images of IPHAS survey \citep{Drew} as well as SDSS and Gaia DR2 surveys. The brightness estimates, obtained from the IPHAS images, are listed in Table \ref{phot2}.

\begin{figure}[ht!]
  \centering
  \includegraphics[width=0.50\textwidth]{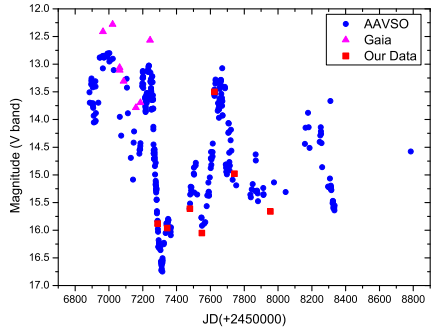}
  \caption{\textit The light curve of V1686~Cyg in V band, with combined data from AAVSO, Gaia and our observations.  \textit{} }
   \label{lightcurve2}
\end{figure}

The important representation of the recent brightness variations of V1686 Cyg give the data from AAVSO (American Associatian of Variable Star Observers), which densely cover the August 2014 - August 2018 period. In Fig. \ref{lightcurve2} we combine the lightcurve in V band from AAVSO with our data and the G magnitudes from Gaia.

Though in 2000-2012 the photometric estimates are scarce, the IPHAS data (Table \ref{phot2}) allow to assume that V1686 Cyg kept its typical behaviour up to 2013, with more or less persistent brightness level near 13.0-13.5 (V). In the first half of 2015 the star demonstrated several characteristic minima. Then, in 2015 Aug the brightness of V1686 Cyg abruptly lowered, reaching V = 16.7 in 2015 Oct. After that the photometric behaviour of the star drastically changed. As one can conclude from the AAVSO lightcurve, at least until the end of 2017 the star remained on the significantly lower level of brightness (about 15.3 in V), with occasional outbursts. During the July-Oct. of 2016 a prominent outburst of at least 2.5 magnitudes in V took place, also traced by our photometry, and which by chance coincided with our spectroscopic observations. One cannot exclude another, lower amplitude brightening in the first half of 2018.

\begin{table}
\caption{V1686 Cyg (LkH$\alpha$ 224) photometry}
\label{phot1}     
\centering        
\begin{tabular}{c c c c c}          
\hline\hline                        
Date & \textit{B} & \textit{V} & \textit{R} & \textit{I} \\    
\hline                     
22.09.2015 & 17.22 & 15.88 & 14.68 & 13.27 \\   
18.11.2015 & 17.18 & 15.96 & 14.87 & 13.56 \\
31.03.2016 & 17.03 & 15.61 & 14.48 & 13.07 \\
10.06.2016 & 17.42 & 16.05 & 14.76 & 13.00 \\
23.08.2016 & 14.82 & 13.50 & 12.48 & 11.58 \\
20.12.2016 & 16.34 & 14.98 & 13.62 & 12.49 \\
20.07.2017 & 16.99 & 15.66 & 14.63 & 13.53 \\
\hline                                       
\end{tabular}
\end{table}

\begin{table}
\caption{V1686~Cyg  brightness on the IPHAS images}
\label{phot2}     
\centering        
\begin{tabular}{c c c}          
\hline\hline                        
Date & \textit{R} & \textit{I} \\   
\hline                     
09.08.2003 & 13.29 & 12.17 \\   
10.08.2003 & 13.29 & 12.15 \\
18.10.2003 & 14.53 & 13.37 \\
11.10.2006 & 14.92 & 13.79 \\
\hline
\end{tabular}
\end{table}

The pronounced spectral variations, observed by us during the 2016 event, are important. Such spectroscopic variability, related to photometric variations, was expected for V1686 Cyg by \citet{Hernandez}.
We describe it in the next section.

\subsection {The spectrum of V1686~Cyg}

\subsubsection {The spectrum of V1686~Cyg in quiescent stage}

As was mentioned above, our spectral observations encompassed the whole period of  the V1686 Cyg outburst. We selected eight spectra of best S/N quality for the further analysis. 

Spectra, taken before the outburst, are quite typical for this star, being similar to the results of \citet{Hillenbrand} and \citet{Magakian-a}. In the red part of spectral range the most conspicuous line is a broad and strong H$\alpha$ emission with superposed weak, blue-shifted absorption feature.   Forbidden emissions of [OI] also are present, though very strong background emission lines make uncertain the estimations of their intensity.  Besides, faint emission lines, mainly belonging to FeII (40) and not described before, were detected. NaI D lines
are not well seen either in emission or absorption; its should be mentioned that previous studies show their pronounced variability. 

After the end of the outburst, in Nov.-Dec. 2016, the spectrum returned to the same appearance, with   H$\alpha$ and several FeII lines in emission; besides, broad-winged H$\beta$  and H$\gamma$ absorptions can be seen in the blue range.
In fact, the upper lines of Balmer series in absorption, though not strong, probably always are present in the 
V1686 Cyg spectrum \citep{Magakian-a,Donehew}. In general, during the time of our observations the spectral type of  V1686 Cyg can be assumed as an early Ae, judging by the broad wings of Balmer absorptions and non-detection of HeI 5876 line (which also shows prominent variability in V1686 Cyg spectrum, according to \citet{Mendigutia}. We did not found any other photospheric absorptions, at least with our spectral resolution.

It should be mentioned that in all spectra we detected one of the strongest  diffuse interstellar bands (DIB) 6284 \AA, also mentioned by \citet{Hernandez}, which is easily distinguished from the nearby atmospheric absorption 6278 \AA. This band   with the similar
intensity was previously detected in the spectrum of the nearby LkH$\alpha$ 225 star \citep{MMAG}. 

\begin{figure}[ht!]
  \centering
  \includegraphics[width=0.48\textwidth]{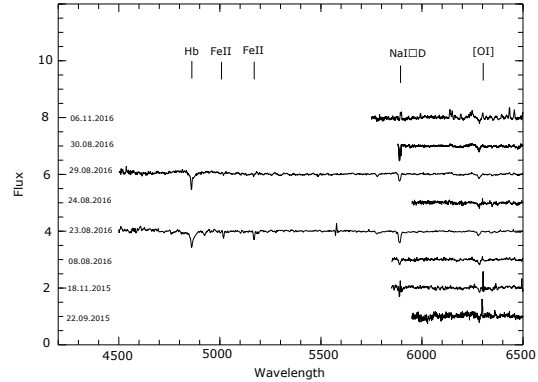}
   \caption{The variations of V1686~Cyg spectrum in 2015-2016, in 4500-6500 \r{A} range. In the shorter wavelengths the H$\beta$ and FeII absorptions are prominent. Also the obvious increase of the chromospheric emission in 6.11.2016 can be seen. }
   \label{minspec}
\end{figure}

\subsubsection{Spectrum of V1686 Cyg during the outburst}

First significant spectral changes one can see in the spectrum of the star, obtained  in  May 16 of 2016, when photometric  variations were not detected yet. The better S/N spectrum of June 10, 2016, still before the photometric brightening, confirms this impression. The central absorption in H$\alpha$  became much stronger, nearly dividing the emission line into two almost equal components. Forbidden emission lines of [OI] and [SII] still are present in the spectrum; their width confirms non-background origin. 

As can be seen from Table \ref{phot1}, the maximal brightness of V1686~Cyg was reached in Aug. 2016. Despite the different spectral resolution, all our five spectrograms of that period have certain similarity. Especially impressive are the further variations in the profile of H$\alpha$ line, where  absorption component lowers beneath continuum.
The strength of iron emissions decreases, forbidden lines of [OI] nearly disappear. On the another hand, rather intensive NaI D absorption lines become visible in the spectrum. In the shorter
wave range, which was observed only with lower resolution,  besides of H$\beta$ and H$\gamma$ with  typical shell cores, we see also absorptions of  FeII (42).  It is worth to mention that after one week these blue iron absorptions disappeared, though the red part of the spectrum kept the same appearance at least until the end of Aug. of 2016. 

\begin{figure}[ht!]
  \centering
  \includegraphics[width=0.5\textwidth]{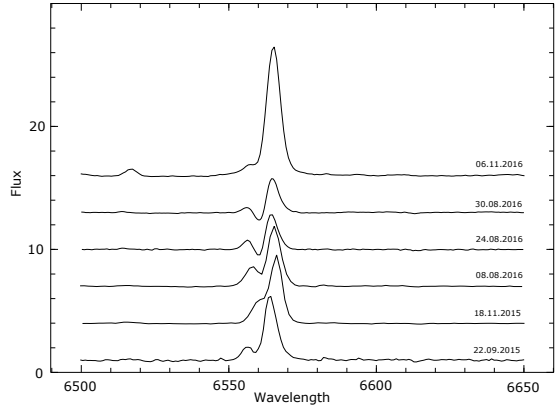}
   \caption{The strong variations in the profile of H$\alpha$ line in the spectrum of V1686~Cyg. During the outburst the absorption component became rather intensive and even lowers below continuum.  }
   \label{maxspec}
\end{figure}

To better represent the changes, we show the parts of the normalised V1685 Cyg spectra (in the range up to H$\alpha$ line),  in various periods in Fig.\ref{minspec}, and the variations in the H$\alpha$ profile are presented in Fig.\ref{maxspec}.

In the 2016 Oct. the star started to lower its brightness (see Fig.\ref{lightcurve}).  In that relation the spectrum, obtained in Nov. 6, is  very interesting. It significantly differs as from the outburst spectra, as well as from the other spectra in quiescent stage. On the one hand, strong NaI D absorptions and   H$\alpha$ line absorption component disappeared. On the other hand, all emission lines became obviously stronger than before the outburst. The emission line spectrum in 6000-6500 \r{A} range appears much more rich, with lines of FeII, FeI and CaI. We can note it remarkable similarity to the spectra of PV Cep and V350 Cep, shown by   \citet{MMAG}. Subsequent spectra, though  of the lower S/N ratio, are more like to the pre-outburst spectra, as it was already mentioned in the previous section.

\subsection{Equivalent widths and radial velocities}

For the better representation of the spectral changes, described above, we estimated the equivalent widths of most typical emission and absorption lines. The general picture seems consistent with the mentioned above conclusion, that the strength of emission lines relative to continuum significantly lowered during the maximal brightness period. For example, EW of H$\alpha$ emission component changed from $\approx -35$ \r{A} to the $\approx -12$ \r{A} in the end of Aug.2016; then high values again  were restored. In the described above spectrum of Nov. 6, 2016 with very strong emissions EW(H$\alpha$) exceeded $-50$ \r{A}. A similar behavior   also  can be seen for  $\lambda 6300 $ \r{A}  [OI] line.
On the other hand, EW of NaI D absorption in the period of the outburst reached 5 \r{A}, while it was actually invisible before.
Pronounced variability of emission as well as absorption lines in V1686 Cyg spectrum also was previously described by \citet{Mora} and by \citet{Hernandez}.

We measured the heliocentric radial velocities of several lines
and found a pattern, very typical for young stellar objects. All measurable absorption lines demonstrated negative velocity. For example, the mean velocity of the absorption component of H$\alpha $ is  $-97 \pm$ 47  km s$^{-1}$. Such large dispersion is natural, because the previous observations also have shown the existence of the blue-shifted absorption components in the H$\alpha $ line profile with various, but always negative radial velocities. The narrow absorption cores in   H$\beta$ and H$\gamma$ have similar velocities, though these values cannot be measured with sufficient accuracy because of the lower spectral resolution.  FeII(42) absorptions, seen in only one lower resolution spectrum, also show ambiguous velocities, probably  because of the same reason. Both lines of NaI D doublet have velocity near $-$57 km s$^{-1}$,  similar negative velocity has also $\lambda$6300 \AA\ [OI] emission.    The peak of  H$\alpha$  emission  has positive radial velocity:  $+95 \pm 29$ km s$^{-1}$.  Velocities of the iron emissions cannot be measured because of  their faintness.
All these data are in sufficiently good compliance with high-resolution measurements, presented by \citet{Cauley}.

 \section{Discussion and conclusions}
     
Our combined data allow to make several conclusions about the observed spectral variability of V1686 Cyg.  These observations fully confirm already described by the listed above authors the pronounced changes in the strength of certain absorption and emission lines,  which easily explain the large range of the spectral types, assigned to this object. Actually, V1686 Cyg is one of the most photometrically and spectrally variable  HAeBe star, and as one can see (at least in this present case), the spectral and photometric variations of  V1686 Cyg are directly related. 

As was already stated above, the observed short-time brightening is not typical for the V1686 Cyg. At least similar events cannot be found on the previous long time range light curve, presented by \citet{Herbst}. it can be considered as an outburst, because its  accompanying spectral changes could be interpreted as the formation of dense expanding envelope around the star, with its subsequent dissipation during several months. This envelope, emitting mainly in continuum, covered up lower layers of the stellar chromosphere, making invisible the metallic emissions and diminishing even very strong emission component of H$\alpha $ line. On the other hand, the envelope was sufficiently dense to produce absorption lines with negative radial velocity.

The permanent existence of the blue-shifted absorption components in the H$\alpha $ line profile, shown by the previous observations \citep{Magakian-a}, allows to conclude, that the similar expanding envelopes, though of much  lower density, nearly always are present around V1686 Cyg star. 

By now we do not know, how long the V1686 Cyg star will remain in its present lower brightness state. Only the new photometric observations can make the situation more clear. This star definitely deserves continuing monitoring.

Several authors make analogies between V1686 Cyg short-time light drops and UX Ori type variability. However, this question remains to be investigated. In fact, this star may be an object, which combines two types of PMS variability, like V2492 Cyg, V350 Cep or V582 Aur \citep{Giannini,Jurdana,Abraham}.

  \begin{acknowledgements}
 This work was supported by the RA MES State Committee of Science, in the frames of the research project number 18T-1C-329.

This research has made use of the VizieR catalogue access tool, CDS, Strasbourg, France (DOI: 10.26093/cds/vizier). This paper makes use of data obtained as part of the INT Photometric H$\alpha$ Survey of the Northern Galactic Plane (IPHAS) carried out at the Isaac Newton Telescope (INT). The INT is operated on the island of La Palma by the Isaac Newton Group in the Spanish Observatorio del Roque de los Muchachos of the Instituto de Astrofisica de Canarias. All IPHAS data are processed by the Cambridge Astronomical Survey Unit, at the Institute of Astronomy in Cambridge.
We acknowledge with thanks the variable star observations from the AAVSO International Database  mainly contributed by observer James McMath, which were used in this research.

The authors are grateful to the referee for valuable suggestions and comments.  
\end{acknowledgements}

\end{document}